\documentclass[aps,prb,superscriptaddress,twocolumn,showpacs, longbibliography]{revtex4-1}

\usepackage{graphicx}
\usepackage{dcolumn}
\usepackage{bm}
\usepackage{scrextend}
\usepackage{hyperref}
\usepackage{tabularx}
\usepackage{amsmath}

\usepackage{filecontents}
\usepackage{color}

\begin{document}


\title{Effect of N interstitial complexes on the electronic properties of GaAs$_{1-x}$N$_{x}$
  alloys from first principles}

\author{Jos\'e D. Querales-Flores}
\email{jose.querales@tyndall.ie}
\affiliation{Tyndall National Institute, Lee Maltings, Dyke Parade, Cork T12 R5CP, Ireland}
\affiliation{Centro At\'omico Bariloche-CNEA and CONICET, Av. Bustillo
  Km. 9.5, 8400-Bariloche, Argentina}
   \affiliation{Instituto Balseiro,  Univ. Nac. de Cuyo and CNEA, 8400-Bariloche, Argentina}
\author{Cecilia I. Ventura}
\affiliation{Centro At\'omico Bariloche-CNEA and CONICET, Av. Bustillo  Km. 9.5, 8400-Bariloche, Argentina} 
  \affiliation{Univ. Nac. de R\'io Negro, 8400-Bariloche, Argentina}
\author{Javier D. Fuhr}
\affiliation{Centro At\'omico Bariloche-CNEA and CONICET, Av. Bustillo
  Km. 9.5, 8400-Bariloche, Argentina} 
  \affiliation{Instituto Balseiro,  Univ. Nac. de Cuyo and CNEA, 8400-Bariloche, Argentina}

\date{\today}

\begin{abstract}
  Although several approaches have been used in the past to
  investigate the impact of nitrogen (N) on the electronic structure
  of GaAs$_{1-x}$N$_{x}$ alloys, there is no agreement between theory
  and experiments about the importance of the different N interstitial
  defects in these alloys, and their nature is still unknown.  Here we
  analyze the impact of five different N defects on the electronic
  structure of GaAs$_{1-x}$N$_{x}$ alloys, using density-functional
  methods: we calculate electronic states, formation energies and
  charge transition levels. The studied defects include
  N$_\textrm{As}$, As$_\textrm{Ga}$, As$_\textrm{Ga}$-N$_\textrm{As}$
  substitutional defects, and (N-N)$_\textrm{As}$,
  (N-As)$_\textrm{As}$ split-interstitial complex defects. Our
  calculated defect formation energies agree with those reported by
  S.B. Zhang et al. [Phys. Rev. Lett. 86, 1789 (2001)], who predicted
  these defects.  Among the interstitial defects, we found that
  (N-As)$_\textrm{As}$ emerges as the lowest energy configuration in
  comparison with (N-N)$_\textrm{As}$, in agreement with recent
  experiments [T. Jen, et al., Appl. Phys. Lett. 107, 221904 (2015)].
  We also calculated the levels induced in the electronic structure
  due to each of these defects: defect states may occur as deep levels
  in the gap, shallow levels close to the band edges, and as levels
  resonant with bulk states.  We find that the largest changes in the
  band structure are produced by an isolated N atom in GaAs, which is
  resonant with the conduction band, exhibiting a strong hybridization
  between N and GaAs states. Deeper levels in the bandgap are obtained
  with (N-N)$_\textrm{As}$ split-interstitial defects. Our results
  confirm the formation of highly localized states around the N sites,
  which is convenient for photovoltaics and photoluminescence applications.
\end{abstract}


\maketitle

\section{Introduction}\label{introduction}

Defects, whether intentional or not, have a dominant impact in both
electrical and optical properties of semiconductor
alloys.\cite{geisz2002} The rich variety of defects along with their
technological importance guarantee interest to their computational
research. III-N-V semiconductors, in particular GaAsN alloys, can be
lattice matched to substrates such as GaAs, Ge and Si, with a range of
direct gaps that are complementary to those of other III-V
semiconductors.\cite{geisz2002,fischer2002,geisz2002,harris2002,kurtz1999,OReillyreview}
There are many applications for these alloys including the
next-generation of high-efficiency multijunction solar cells,
high-performance electronic devices and long-wavelength light emitters
and detectors.

 Intrinsic point defects have been identified in GaAsN as: As$_\textrm{Ga}$ antisites\cite{tkinh2001}, N interstitials\cite{spruytte2001,ahlgren2002} and Ga vacancies\cite{wei2001}.
  Experimental and theoretical works have suggested that the majority of
N atoms incorporated to GaAs could be expected to go to isoelectronic
substitutional As sites.\cite{zhang2001,carrier2005}However, Zhang and
Wei\cite{zhang2001} suggested that a significant fraction of N atoms
incorporate non-substitutionally as split interstitials,
(N-N)$_\textrm{As}$ or (N-As)$_\textrm{As}$.
For substitutional N atoms on As sites, it is energetically more favorable to be somewhat apart from each other and thus the equilibrium thermodynamics does not favor the formation of nitrogen pairs\cite{virkkala2012}. The reason for this is that although the total energy can be lower when two defects share their surrounding shear fields, two N atoms being too near to each other causes larger structural changes into the host crystal than being somewhat more apart. In dilute N containing GaAs, it has been reported that the presence of (N$_\textrm{As}$-N$_\textrm{As}$) substitutional pairs is negligible\cite{ivanova2010}.
  Direct measurements of
the fraction of N incorporated interstitially in GaAs via nuclear
reaction analysis (NRA) have been reported by several
groups. \cite{beaudry2004,reason2004,ahlgren2002,spruytte2001,jen2015}
More recently, Jen et al.\cite{jen2015} reported a comparison between
Rutherford backscattering spectroscopy and NRA with Monte
Carlo-Molecular Dynamics simulations, in order to distinguish
(N-N)$_\textrm{As}$, (N-As)$_\textrm{As}$ and
(As$_\textrm{Ga}$-N$_\textrm{As}$) complexes in GaAsN alloys. The
results suggested that (N-As)$_\textrm{As}$ is the dominant
interstitial complex in dilute GaAsN, in contrast to previous
electronic structure calculations, based on density functional theory
(DFT), which predicted (N-N)$_\textrm{As}$ to be the most
energetically favorable configuration.\cite{laaksonen2008,carrier2005}
In 2017, Occena et al.\cite{occena2017} examined the influence of the
flux on bismuth and nitrogen incorporation during molecular beam
epitaxy of GaAs$_{1-x-y}$N$_{x}$Bi$_{y}$ alloys, and observed an
enhancement in total N incorporation via the formation of additional
(N-As)$_\textrm{As}$.
It is worth mentioning that the presence of split interstitials have been reported in other
III-V semiconductors, for instance: (N-As) bonding defects in (Ga,In)(N,As)/Ga(N,As) heterostructures\cite{ishikawa2011}, and  TlInGaAsN alloys\cite{kim2013}; (Sb-N) split interstitials in (InGa)(AsSbN)/GaAs quantum wells\cite{chen2010} and 
 InSb$_{1-x}$N$_{x}$ alloys\cite{hari2012,pham2007}.

In Ref. \onlinecite{virkkala2012} the impact of  (N-N)$_\textrm{As}$ interstitial, N$_\textrm{As}$-N$_\textrm{As}$ substitutional pairs and N-substitutional on the electronic properties of GaAsN alloys was studied within  hybrid functional scheme (HSE06). However,  the relationship between N incorporation as (N-As)$_\textrm{As}$  and (As$_\textrm{Ga}$-N$_\textrm{As}$) has not been reported yet. Nevertheless, several theoretical approaches have been used in order to investigate the impact of N on the electronic structure of
GaAs$_{1-x}$N$_{x}$ alloys.  For instance,
first-principles,\cite{wang2001,agrawal1998,laaksonen2008,carrier2005}
empirical pseudopotential,\cite{mattila1999,kent2001} and
$sp^{3}s^{*}$
tight-binding\cite{Gladysiewicz2013,lindsay1999,lindsay2001}
calculations of supercells have been performed, giving valuable
insight on the microscopic mechanisms of formation of band-edge states
in GaAs$_{1-x}$N$_{x}$.  The band structure of GaAs$_{1-x}$N$_{x}$
substitutional alloys was studied in the framework of DFT within the
hybrid functional scheme (HSE06),\cite{virkkala2012} and it was found
that the trends in the bandgap reduction in these alloys result mainly
from the positions of the N-induced states with respect to the bottom
of the bulk conduction bands.  Ref. \onlinecite{Wilkins2010}
demonstrated the accuracy of HSE06 hybrid functional for computing the
band offsets of semiconductor alloy heterostructures, including GaAs.
The formation energies and transition levels of a set of
interstitial-type defects of N in GaAs have been determined by DFT
calculations, in local density approximation (LDA)\cite{laaksonen2008}
and generalized gradient approximation (GGA).\cite{carrier2005}
Nevertheless, for N-N split interstitial defects, differences were
observed between the bandgap transition levels calculated with LDA and
GGA approximations.\cite{laaksonen2008,carrier2005}

It is known that point-defect formation energies calculated within the
framework of density functional theory often depend on the choice of
the exchange and correlation ($xc$) functional.  Regarding that,
Freysoldt, Lange and Neugebauer\cite{freysoldt2016} showed that
variations between the LDA, GGA, and hybrid functionals emerge from
differences in the position of the bulk valence-band maximum, as well
as in the reference energies for the chemical potential obtained with
distinct $xc$ functionals. Freysoldt et al.\cite{freysoldt2016} used a
band-alignment strategy, based on aligning a benchmark defect level
between GGA and hybrid functional calculations; they also included
changes to reference energies for chemical potentials between
functionals.

In this work, we determine formation energies and charge transition
levels for a large set of N defects in GaAs using a supercell approach
for defects calculations in GGA aproximation and hybrid functionals
calculations for the reference energies for the chemical potential,
adopting the framework recently proposed in
Ref.\onlinecite{freysoldt2016}.  In particular, we investigate the
case of single-atom substitutional defects N$_\textrm{As}$ and
As$_\textrm{Ga}$. We also consider complex defects such as
As$_\textrm{Ga}$-N$_\textrm{As}$, in which the N$_\textrm{As}$ defect
binds to one As$_\textrm{Ga}$ antisite, and split interstitials
N-N$_\textrm{As}$ and N-As$_\textrm{As}$, in which a dimer of N-N or
N-As is located in an As site\footnote{\protect{The effects of hydrogen on the electronic properties of dilute GaAsN Alloys have been studied in Ref.\onlinecite{janotti2002}, where the results showed that nitrogen alloying qualitatively alters the electronic behavior of monoatomic hydrogen to become only a donor, despite the fact that in GaAs or GaN hydrogen is amphoteric, existing in either donor or acceptor states. At higher hydrogen concentrations, it is energetically more favorable to form self-compensated, charge neutral H*2N complexes. The formation of these complexes completely eliminates the band gap reduction caused by nitrogen.}}.  Our results suggest that (N-As)$_{As}$
is the dominant split interstitial in dilute GaAsN, in agreement with
recent experimental results.\cite{jen2015,occena2017} We also
calculated the induced levels in the electronic structure due to each
of these defects. Depending on the defect, these states occur as deep
levels in the band gap, as shallow levels very close to the band
edges, as well as levels in-between the bulk states.  We find that the
most prominent of these levels is due to an isolated N atom in GaAs,
which is resonant with the conduction band, exhibiting a strong
hybridization between N and GaAs states. Deeper levels in the energy
gap have been found for N pairs. These results confirm the formation
of highly localized electronic states around the nitrogen sites, which
is favourable for applications.\cite{shafi2009,OReillyreview,tan2013} In addition, our results for the  effect of the substitutional N$_{As}$ defect on the electronic structure of GaAs agree very well with those given by the well-known band anticrossing model\cite{walukiewicz2002}, therefore providing further support to it.

This paper is organized as follows. In Sec. \ref{approach}, we
describe our theoretical approach based on DFT calculations. In
particular, we discuss the method we adopted to calculate the defect
formation energies by combining GGA and hybrid functionals
calculations. The results of our calculations are given in
Sec. \ref{results}, where they are compared to previous theoretical
and experimental results. In Sec. \ref{conclusions}, the conclusions
are summarized.

\section{Theoretical approach}\label{approach}

The most relevant thermodynamic quantity characterizing a point defect
$X$ in charge state $q$ is its formation energy, as comprenhensively
reviewed by Van de Walle, Neugebauer et al. in
Refs. \onlinecite{vandewalle2004},\onlinecite{freysoldt2014b}, given
by:
\begin{equation}\label{formationenergy}
  E_{f}(X^{q}) = E_{t}[X^{q}] - E_{t}[Bulk] - \sum_{i}{n_{i}\mu_{i}} + q[\mu_{e} + \Delta V] 
\end{equation}
which depends on the chemical potentials $\mu_i$ of atoms that have
been added ($n_{i}>0$) or removed ($n_{i}<0$) and the chemical
potential for electrons $\mu_{e} = E_{F} + E_{VBM}$, being $ E_{F}$
and $E_{VBM}$ the Fermi level and the valence band maximum
respectively. $E_{t}[X^{q}]$ and $ E_{t}[Bulk]$ are the calculated
total energies of the defect $X$ (in charge state $q$) and bulk
respectively. The chemical potentials $\mu_{i}$ of the $n_{i}$ added
or removed atoms allow us to take into account the growth
conditions. $\Delta V$ corresponds to a correction for the finite size
of charged supercells, in order to align the electrostatic
potentials.\cite{laaksonen2008} In our present work, $\Delta V$ was calculated as the difference of the electrostatic potential in the host and defects supercells far from the defect\cite{zunger1998}.

We have determined the formation energies and transition levels of
nitrogen defects in GaAs, with spin-polarized DFT total energy
calculations. All defect configurations are obtained through full
structural relaxation carried out within a DFT framework in which the
exchange-correlation energy is described through the semilocal
approximation proposed by Perdew, Burke, and
Ernzerhof (PBE). \cite{perdew1996,perdew1997} To obtain the chemical
potential of atoms, $\mu_i$ in Eq.\ref{formationenergy}, we performed
calculations using the screened hybrid functional of Heyd, Scuseria,
and Ernzerhof (HSE).\cite{hse1,hse2} We used the VASP (Vienna
Ab-initio Simulation Package) code\cite{vasp} with the projector
augmented wave (PAW). We used a plane-basis set defined by a kinetic
energy cutoff of 400 eV and the gallium $3d$ states were treated as
valence states. We performed calculations using 64 and 65-atoms
supercells with a 3$\times$3$\times$3 $k$-point set generated using
the Monkhorst-Pack method.  We study the convergence of the total energy 
with respect to the $\vec{k}$-point grid: we got converged results for a sampling of 3$\times$3$\times$3.
We have used a supercell size as typically used in previous studies\cite{laaksonen2008,virkkala2012}. In Ref. \onlinecite{laaksonen2008},  64 and 216-atom supercells gave similar formation energies (less than 70 meV) and transition levels showing good convergence as a function of the supercell size.

For neutral defects the hybrid functional calculation were incorporated as a correction for the chemical potentials of the individual species in Eq. \ref{formationenergy}, in our case Ga, As and N. As shown in Ref. \onlinecite{freysoldt2016}, for Ga and As. For calculating corrections we require that the corrections reproduce the HSE enthalpy of formation ($\Delta H^{f,HSE}$)  of N$_\textrm{2}$, GaN and GaAs, by taking HSE as the best available theory reference. We obtained PBE corrections for Ga, As and N: +0.075 eV, -0.689 eV and +0.68 eV, respectively. As discussed in Ref. \onlinecite{freysoldt2016}, when appropriate corrections are applied, the standard local and semilocal functionals can be used to screen for relevant defect configurations before using the computationally more demanding as HSE hybrid functionals or GW.

On the other hand, for charged defects the chemical potentials for electrons is needed. We corrected the position of the bulk valence band entering Eq. \ref{formationenergy} as reference for the Fermi level\cite{freysoldt2016}. 
Ref. \onlinecite{freysoldt2016} shows that this approach corrects the known weakness of a functional in describing the extended host states, while leaving the description of the localized defect states unaltered. This is appropriate for defects that possess well-localized defect states within the band gap.

  We fully relax the unit cell to obtain the
lattice constants for GaAs of 5.7497 \, \AA, consistent with earlier
reports\cite{colleoni2016}. The atomic positions are relaxed until the
Hellmann-Feynman force acting on each atom is reduced down to less
than 1.0 meV/\AA. 

The unfolding of the supercell bandstructure has
been performed using the BandUP code\cite{madeiros2014,madeiros2015}.
In order to have a good description of the band gaps, for the calculation of electronic properties (density of states and band structure) we used the modified Becke-Johnson (mBJ) exchange potential in combination with LDA-correlation\cite{Becke2006,Tran2009}, and GGA structural parameters.

\section{Results and discussion}\label{results}
  
Here, we will present structural and electronic structure results
obtained using our approach described in Sec.\ref{approach}, in four
subsections.
As mentioned in Introduction, the substitutional N$_\textrm{As}$ defect has
been extensively studied in comparison with those interstitials ones.
In our analysis of the electronic structure of GaAsN alloys, we will
always include our own results for N$_\textrm{As}$ as a validation of the approach we used.
In Sec. \ref{structural}, we will discuss and compare our results for
the structural properties for the N defects studied, with recent
theoretical values, when available. In Sec. \ref{dos}, we show our
results for the total and projected densities of states, in order to
identify the N-induced electron states around the gap of GaAs. In
Sec.\ref{transitions} we analize the formation energies for charged
defects, for all studied N configurations as a function of Fermi
energy. In Sec.\ref{unfolding}, we will describe the unfolded band
structure for each N defect in connection with the results first
presented in Sec.\ref{dos}.
  
\subsection{Structure of the studied defects}\label{structural}

\begin{figure}[t!]
\centering
  \begin{tabular}{cc}
   (a)   \includegraphics[width=.35\columnwidth]{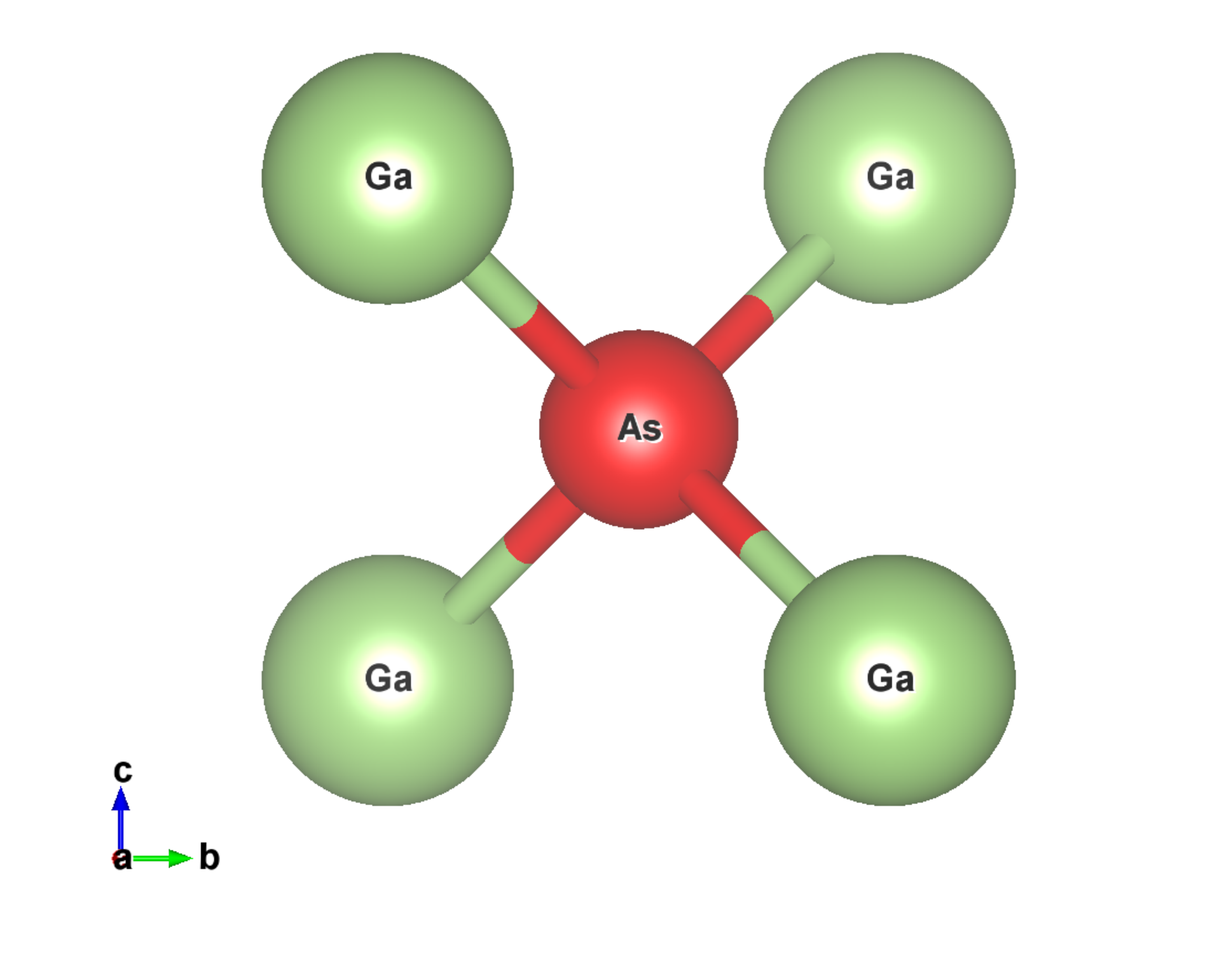} &
  (b)   \includegraphics[width=.35\columnwidth]{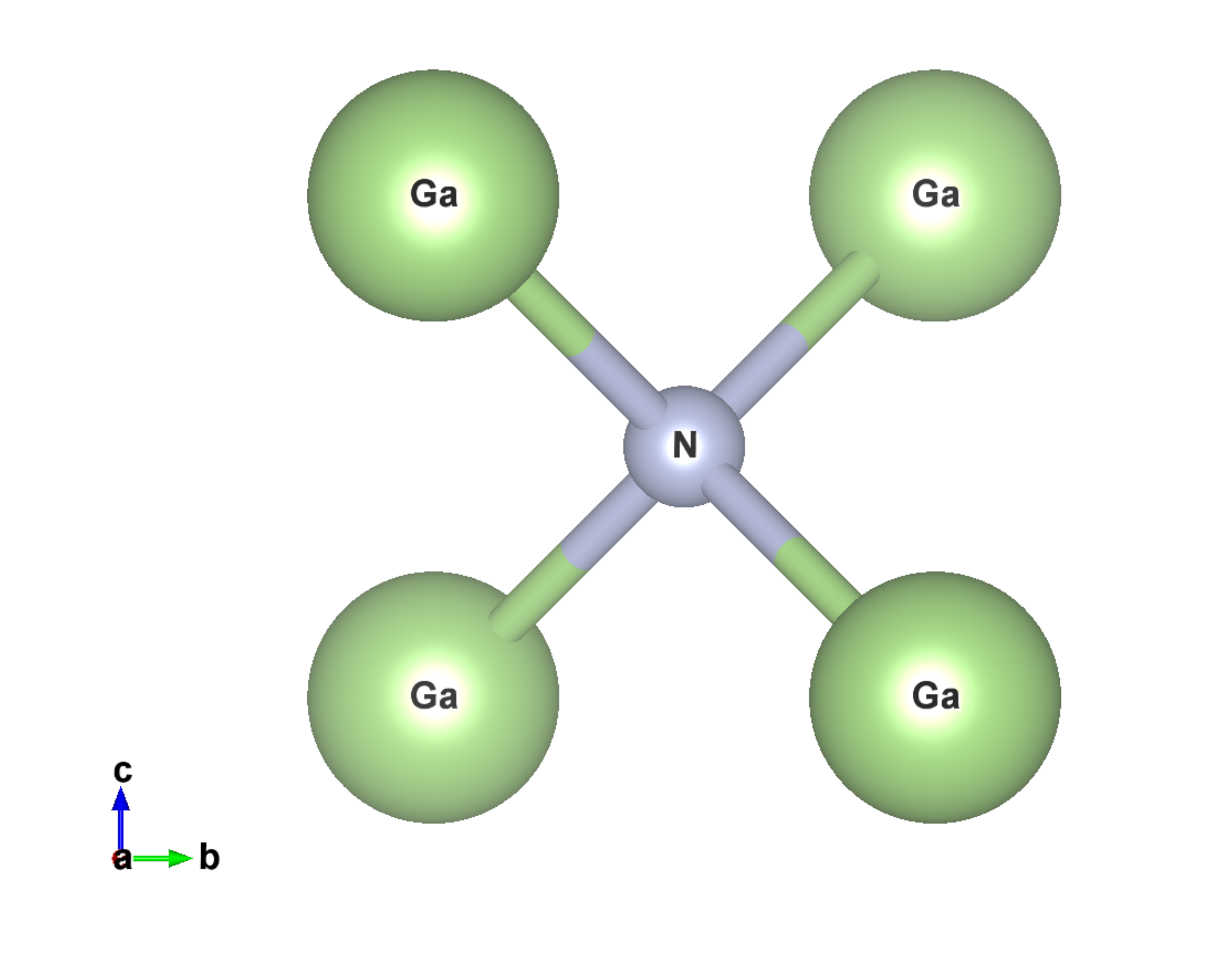} \\
  (c)   \includegraphics[width=.35\columnwidth]{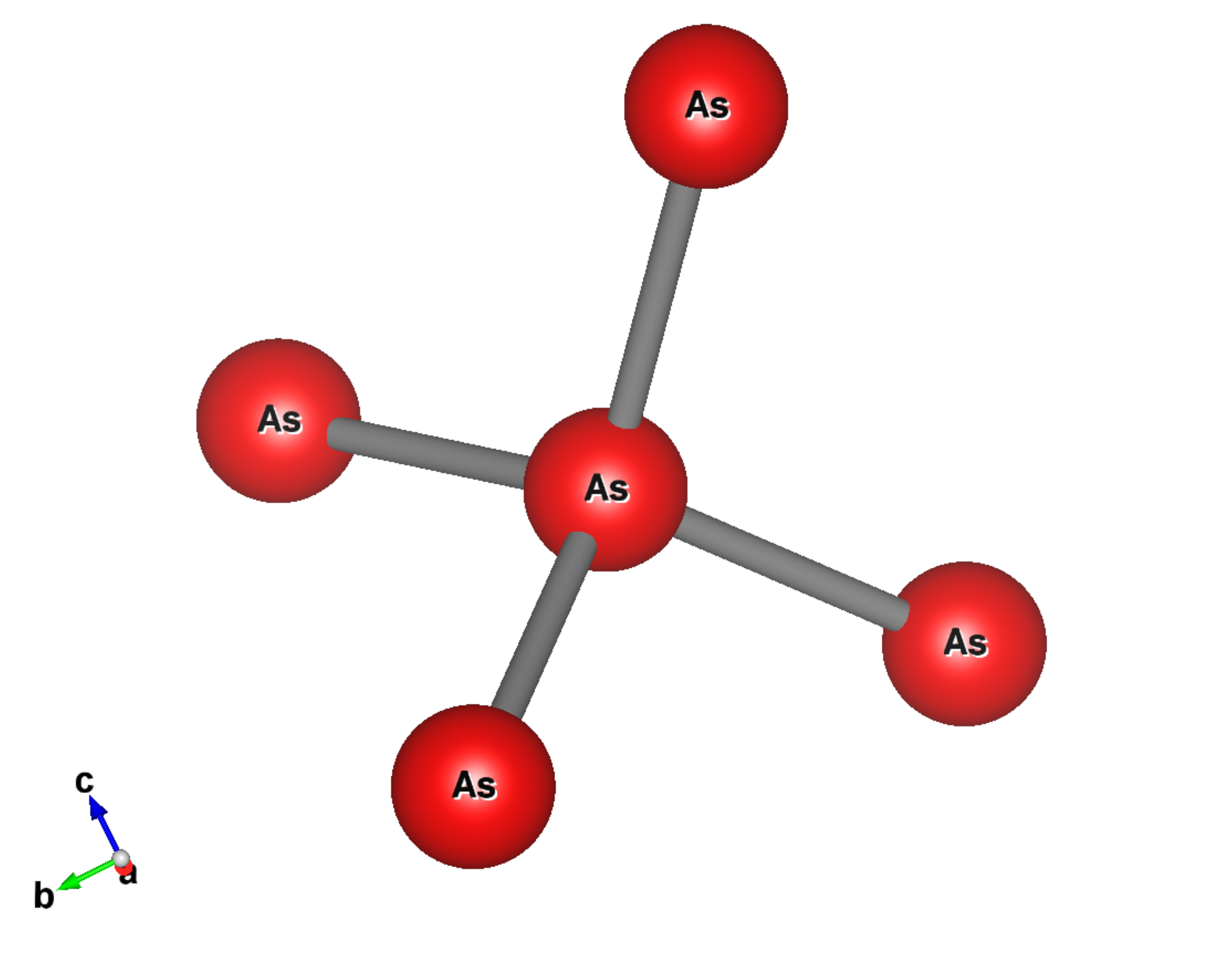}  &
 (d)    \includegraphics[width=.35\columnwidth]{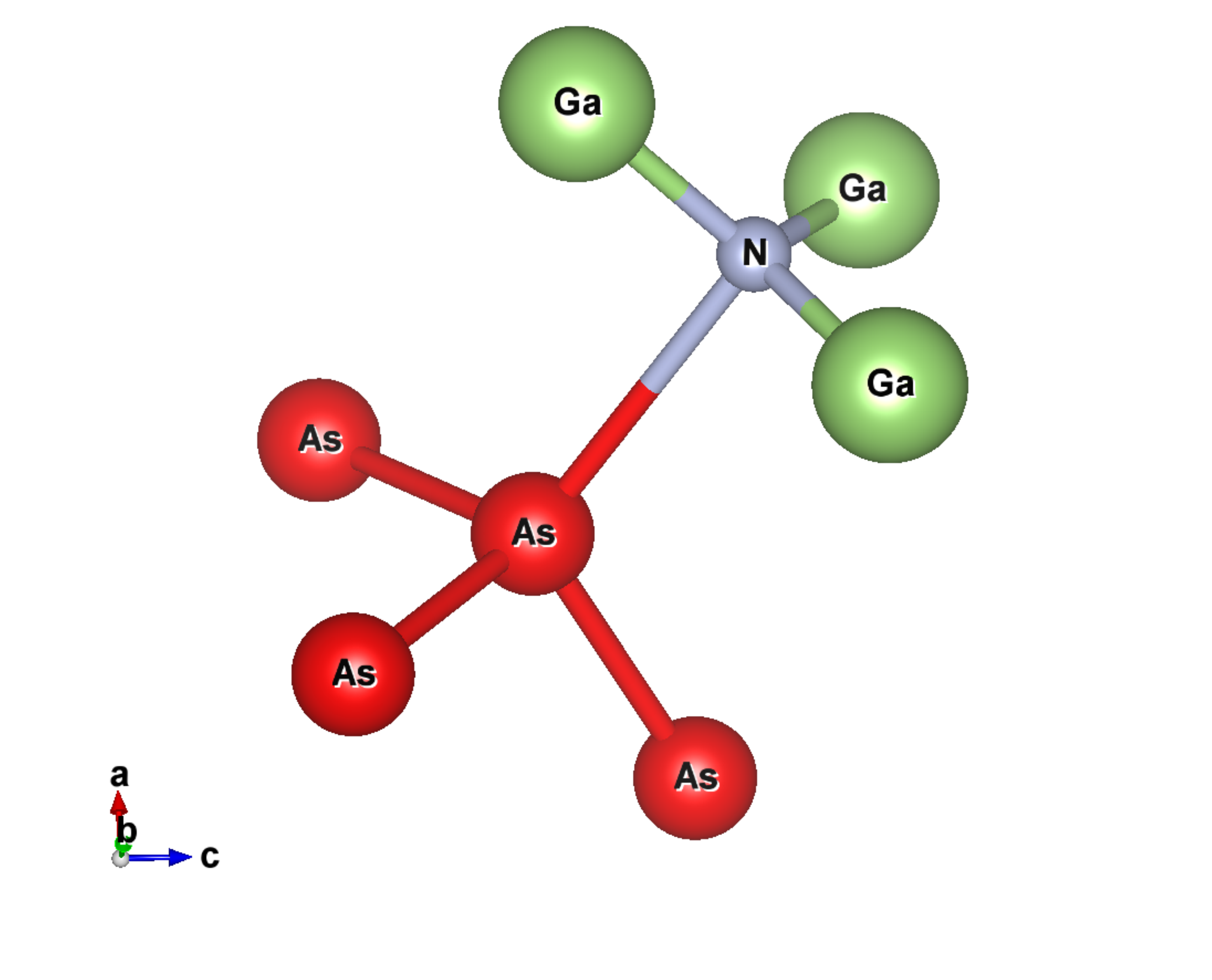} \\
 (e)    \includegraphics[width=.35\columnwidth]{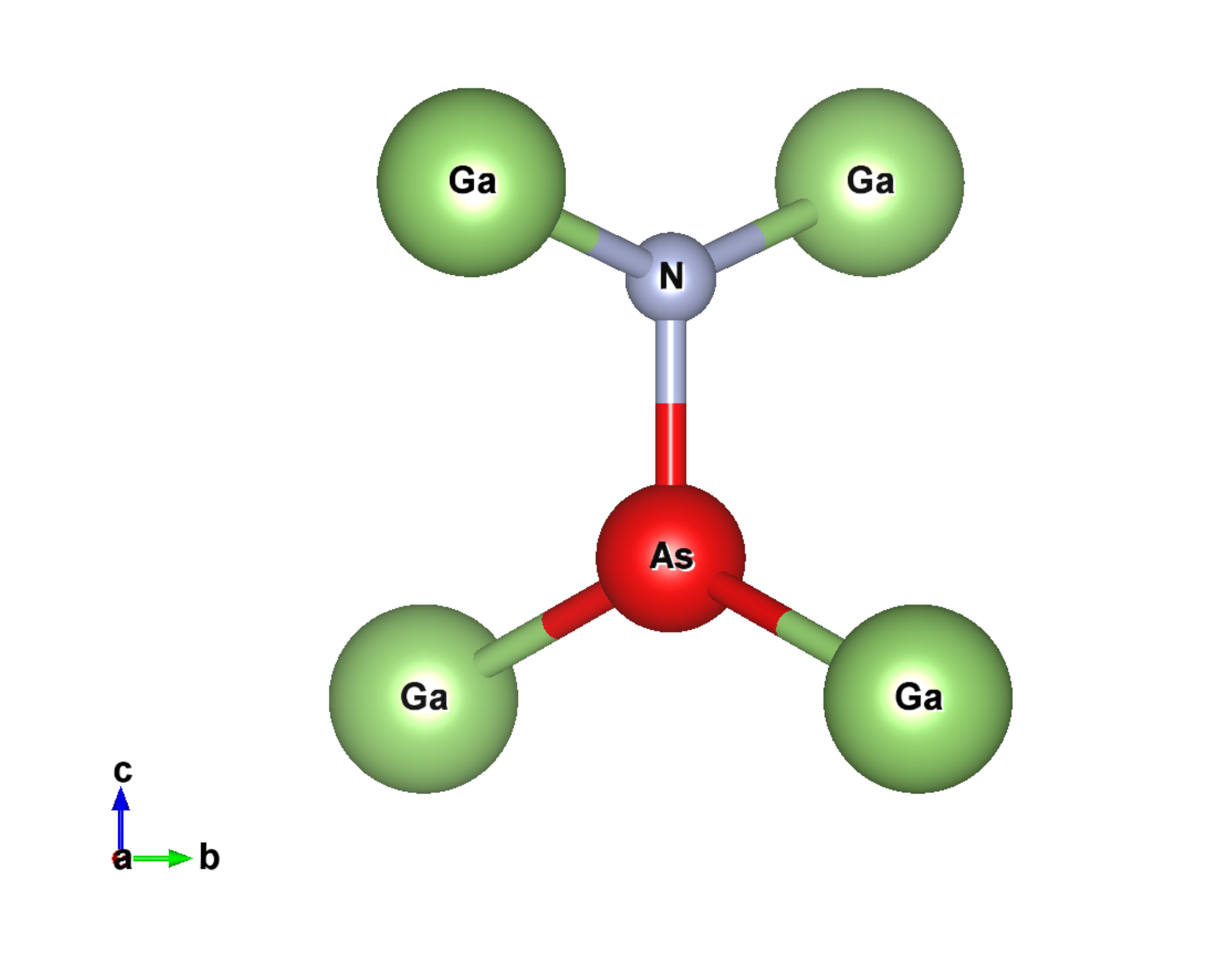}    &
 (f)    \includegraphics[width=.35\columnwidth]{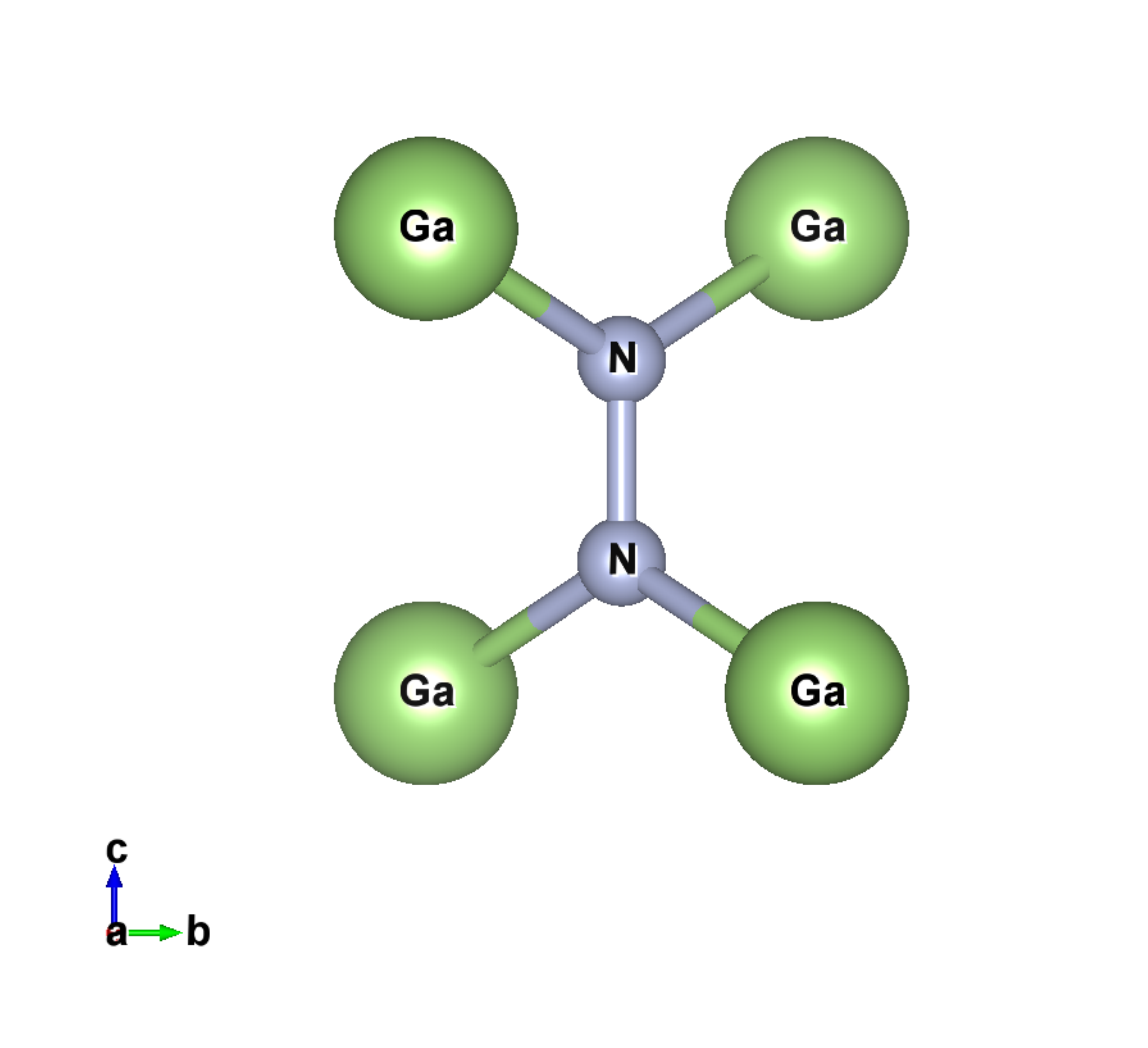}
     \end{tabular}
     \caption{Configurations of native defects in
       GaAs$_{1-x}$N$_{x}$. (a) Crystalline GaAs, (b) single N atom
       located in an arsenic site, N$_\textrm{As}$, (c) single arsenic
       atom located in a gallium site, As$_\textrm{Ga}$, (d)
       N$_\textrm{As}$ defect binds to one As$_\textrm{Ga}$ antisite,
       As$_\textrm{Ga}$-N$_\textrm{As}$, (e) N-As pair in an arsenic
       site, (N-As)$_\textrm{As}$ and (f) N-N dimer in an arsenic
       site, (N-N)$_\textrm{As}$. The figures were generated using
       VESTA software.\cite{vesta}}\label{defects}
\end{figure}   

Figure \ref{defects} shows the structure of an As atom and its
neighbours in GaAs (Fig. \ref{defects}(a)), along with the five
complex defects studied in this work (Figs. \ref{defects}(b)-(e)).
When a N atom is incorporated in an As substitutional lattice site,
the following bonds are formed with respective bond-lengths:
2.1119~\AA\ for N-Ga, and 2.4902~\AA\ for As-Ga.  We found that N
incorporates in the split interstitial (N-N), in which two N atoms
share a substitutional lattice site with a strong bond between them:
1.3349 \AA \, is the N-N bond-length and 2.0220 \AA \, the N-Ga
one. The N-N bond-length results are consistent with previous LDA
studies: i.e.  1.34 \AA\cite{laaksonen2008}, and 1.33
\AA\cite{carrier2005}; while the value obtained for the N-Ga
bond-length is slightly larger than 1.95 \AA\, reported in
Ref.~\onlinecite{laaksonen2008} for LDA.

When N incorporates in the other split interstitial configuration, in
which a dimer with N-As atoms occupies a substitutional lattice site
and we find 1.7998 \AA\, for N-As bond-length (LDA:
1.79\AA\,\cite{laaksonen2008}), 2.4351 \AA\, for As-Ga bonds (LDA:
2.38 \AA\, \cite{laaksonen2008}) and 1.9298 \AA\, for N-Ga bonds (LDA:
1.88 \AA\, \cite{laaksonen2008}).  For substitutional As$_\textrm{Ga}$, we
find that 2.6085 \AA\, As-As, and 4.0893 \AA\, As-Ga bonds are formed.

\begin{figure}[h]
  \begin{center}
    \includegraphics[width=0.9\columnwidth]{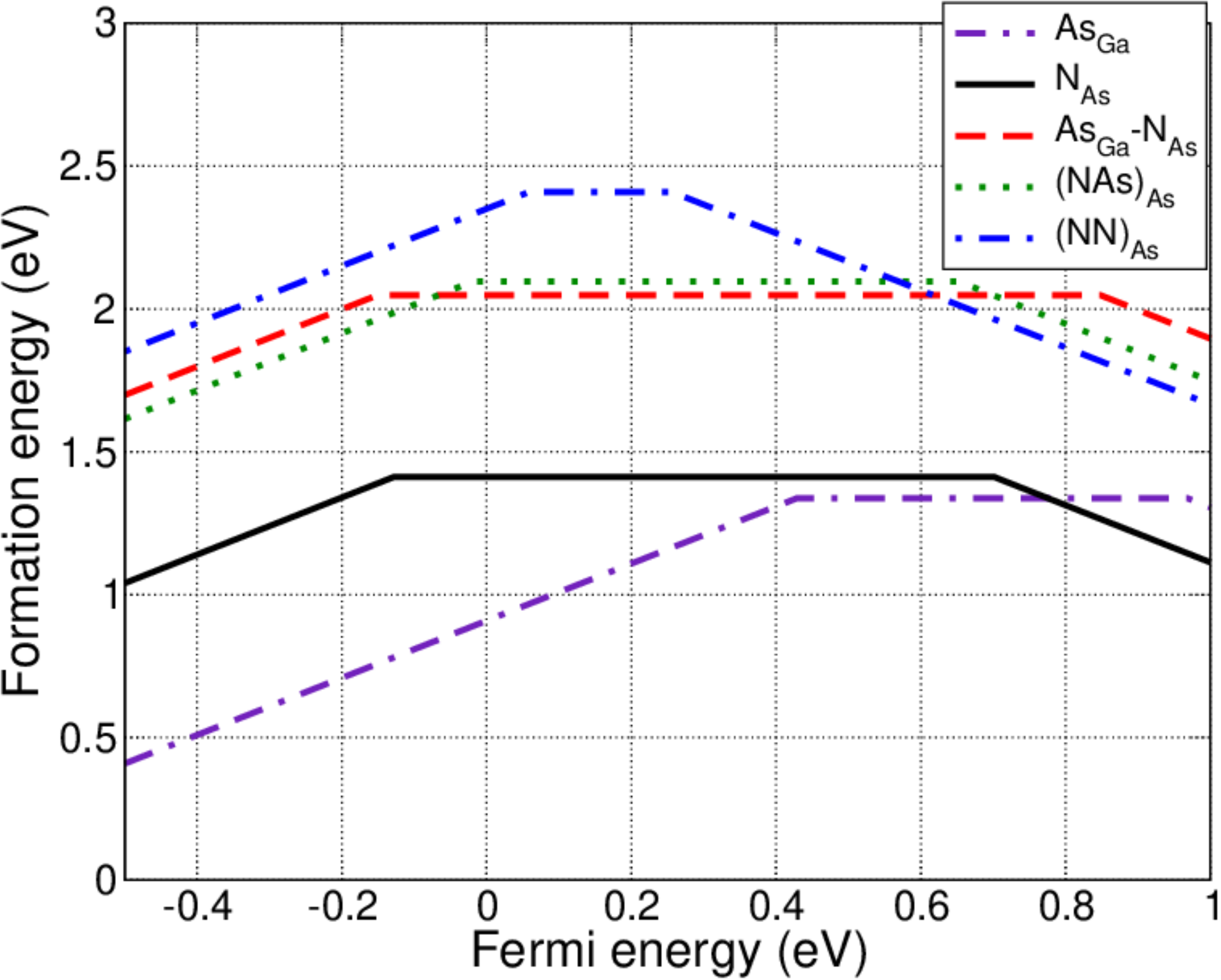}
    \caption[]{Formation energies vs. Fermi level for substitutional,
      interstitial, and antisite defects in GaAs$_{1-x}$N$_{x}$,
      calculated with a GGA and HSE hybrid functionals corrections, as
      proposed by Freysoldt et al. in Ref. \onlinecite{freysoldt2016}.
      All sets of formation energies include finite-size corrections
      (see last term in Eq. \ref{formationenergy}). The Fermi-level
      range is extended beyond the band maxima. In all cases,  As-rich
      conditions are chosen: i.e. $\mu_{N}=\mu_{N_{2}/2}$. The lattice constant was taken as the
      DFT-GGA one, mentioned in Sec.\ref{approach}.}.
    \label{estability}
  \end{center}
\end{figure}

\subsection{Thermodynamic transition levels of N defects in
  GaAs$_{1-x}$N$_{x}$}\label{transitions}

According to Eq.~(\ref{formationenergy}) the higher the Fermi level
is, the more energy is gained by moving the electron from the
reservoir into the defect.  That is, the formation energy of a
negatively charged defect gets lower at higher values of the Fermi
level, and viceversa for the positively charged defects. If the defect
has the lowest formation energy in the negative charge state, then a
neutral defect would like to receive an electron from the electron
reservoir, so that it behaves like an acceptor. In a similar way, if a
neutral defect likes to donate an electron to the electron reservoir,
it becomes positively charged and this can be related with a donor.
When the formation energies of all the possible charge states are
calculated, the charge state yielding the lowest formation energy
depends on the Fermi level. Consequently, there will be a value of
$E_F$ where the formation energy is the same for two charge states,
which is usually denoted as ionization level, hereafter denoted as
$\epsilon(q_{1}/q_{2})$. This level is the same as the energy required
to promote an electron from the valence band into the defect for
acceptor levels or demoting an electron from the defect to the valence
band for donor levels. The atomic structures of the defects in charge
states $q_1$ and $q_2$ can be different. If the atomic relaxation
occurs during the ionization process, it is called thermodynamic
transition. If not, it is called an optical transition. Formation
energies for thermodynamic transitions of native defects in
GaAs$_{1-x}$N$_{x}$ calculated with GGA (PBE) and HSE are presented in
Fig. \ref{estability}.   As-rich
      conditions in Eq. \ref{formationenergy} have been chosen: i.e. $\mu_{N}=\mu_{N_{2}/2}$ . In
our calculation, for each charged defect configuration a fully
structural relaxation was done.
  

\begin{table}[ht]
  \centering
  \caption{ Charge transition levels ($\epsilon_{q/q'}$), Fermi energy
    for neutral defects ($E^{0}_{F}$) and formation energies ($E_{f}$)
    calculated at the PBE level, and including hybrid functional
    corrections for electron and chemical reservoirs and Ga 3d states
    in the valence band, for the most relevant defects. For
    comparison, we also show corresponding formation energies
    calculated in
    Refs.\onlinecite{laaksonen2008,carrier2005,colleoni2016,orellana2001}.
    $E^{0}_{F}$ for GaAs: 2.717 eV.  }
  \begin{tabular}{ | c | c | c | c | c| }
    \hline
    Defect &  $E^{0}_{F}$ (eV)  &  $E_{f}$ (eV)  &  $\epsilon_{+1/0}$ (eV)  &  $\epsilon_{0/-1}$ (eV)  \\
    \hline
    As$_\textrm{Ga}$ & 3.280  & 1.337  & 0.428  & 0.969    \\
           &   & (1.33\cite{colleoni2016}) &   &    \\ \hline
    N$_\textrm{As}$ & 2.633 &1.406      & -0.128  & 0.7  \\    
           &  & (1.087\cite{carrier2005})      &   &  (0.82 \cite{orellana2001}) \\  \hline  
    As$_\textrm{Ga}$-N$_\textrm{As}$ & 2.635  &2.048   &  -0.151 & 0.848  \\ \hline 
    (N-As)$_\textrm{As}$ & 3.118 &2.178    &  -0.018 &  0.651    \\
    &  & (2\cite{laaksonen2008})   &   &   (0.29\cite{laaksonen2008})   \\\hline 

(N-N)$_\textrm{As}$ & 2.730 &2.462   & 0.058  & 0.256     \\
 &  & (2.754\cite{carrier2005})  & (0.2\cite{carrier2005}) &  (0.12\cite{laaksonen2008}, 0.3\cite{carrier2005})    \\\hline 
  \end{tabular}
  \label{formationenergies}
\end{table}

Figure \ref{estability} shows the calculated formation energies for
all studied defects as a function of Fermi energy. In principle the Fermi level ranges from the VBM up to CBM. We chose the range of Fermi level to be $\sim$1 eV above the VBM based on the location of the defects states identified (it will be discussed in Section \ref{dos}). In the following
we describe the main results.

\textit{N-N split interstitials}: N gives rise to two transition
levels within the band gap of GaAs$_{1-x}$N$_{x}$. Concretely, the
transition from the neutral to $+1$ charge state (0/+1) happens when
the Fermi energy is 0.058 eV above the valence band maximum, while the
transition from the neutral to $-1$ charge state (0/-1) occurs at a
Fermi energy equal to 0.256 eV.  This defect was found to be stable in
positively charged states when the Fermi level is in the lower part of
the band gap, stable in the neutrally charged state for a limited
number of values of the Fermi level in the gap, and becoming
negatively charged at higher Fermi levels.

\textit{N-As split interstitials}: Only one transition level within
the band gap of GaAs$_{1-x}$N$_{x}$ is found, from neutral to $-1$
charge state (0/-1) at the Fermi energy 0.651 eV. For most Fermi
energy values the interstitial defect with the lowest energy is
(N-As)$_\textrm{As}$.
 
\textit{N substitutional}: We find one transition level in the band
gap for the substitutional N$_\textrm{As}$ and As$_\textrm{Ga}$-N$_\textrm{As}$ antisite
defects: the transition (0/-1) happens for the $N_\textrm{As}$ and
As$_\textrm{Ga}$-N$_\textrm{As}$ defects at Fermi energies of 0.7 and 0.848 eV above
VBM, respectively. It is worth mentioning that in previous LDA and GGA calculations in Refs. \onlinecite{laaksonen2008,carrier2005}, no such charged states transitions for  N$_{As}$ had been found.
However, LDA calculations reported in Ref.\onlinecite{orellana2001} the 0/-1 transition was predicted 
to occur at Fermi level $\sim$0.82 eV above the maximum of the valence band.
We did not find experimental studies on it.

\textit{As$_\textrm{Ga}$ substitutional}: we find two transition levels in
the band gap: (+1/0) happens defect at Fermi energy 0.428 and (0/-1)
at 0.969 eV above VBM, respectively.  The occupation of the Ga site by
As leads to insignificant lattice distortions, involving the formation
of defects with the lowest formation energy.

The calculated charge transition levels, Fermi energy for neutral
defects and formation energies, are listed in Table
\ref{formationenergies} together with the results of previous
first-principles calculations, for comparison.

\subsection{Electron density of states}\label{dos}

The GaAs direct bandgap $E_{g}$ obtained with GGA is much smaller with
respect to the experimentally reported one ($\sim
1.52$eV\cite{Wilkins2010,Chtourou2002}): using the GGA optimized
lattice parameter $E_{g} = 0.18$~ eV, which increases to 0.56~eV if
the experimental lattice parameter is used in the calculation. For a
better description of the electronic structure of bulk GaAs and its
defects, we used the mBJ exchange potential in combination with
LDA-correlation\cite{Becke2006,Tran2009}, which is known to yield band
gaps with an accuracy similar to hybrid functional or GW methods. With
this approximation the calculated bulk GaAs band gap is $\sim$1.64 eV,
much closer to the experimental value.

\begin{figure}[h]
  \begin{center}
 \includegraphics[width=0.9\columnwidth]{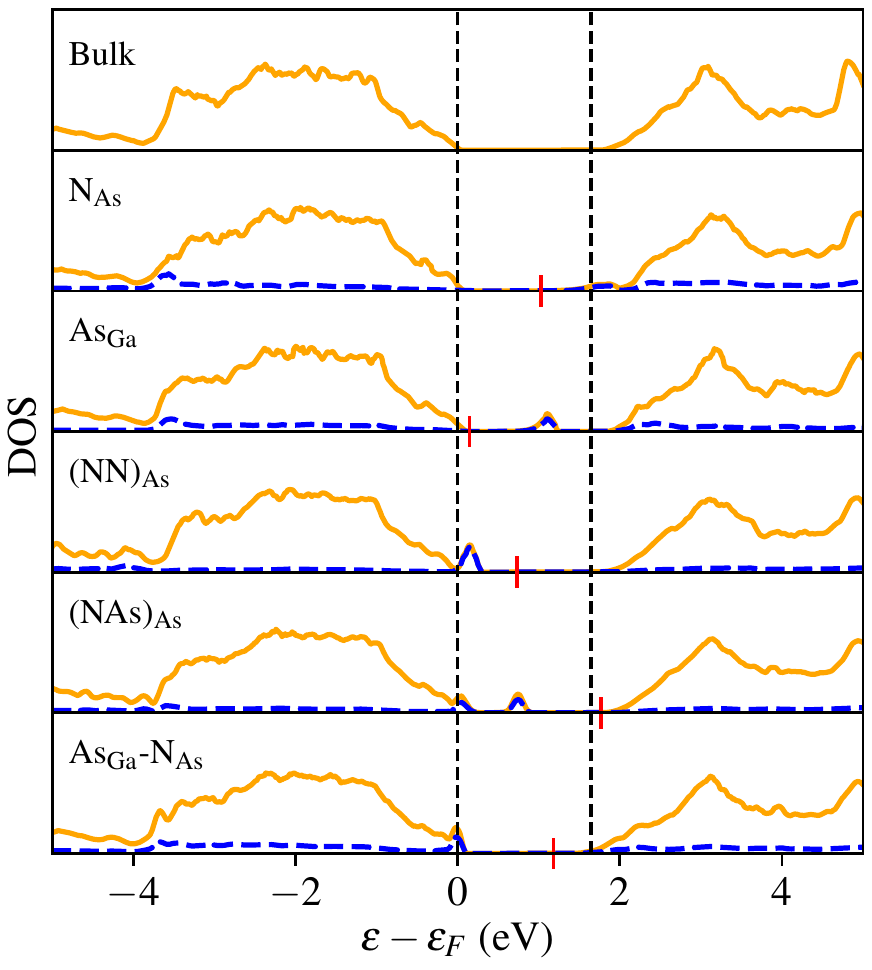}
 \caption[]{Total density of states (DOS) for bulk GaAs and the five
   defect configurations. Vertical lines along the whole figure
   indicate the calculated band gap for pure GaAs. Full lines
   represent the DOS and dashed lines the projected DOS on each defect
   including its first neighbours. The PDOS corresponding to
   $N_\textrm{As}$ was multiplied by a factor of 10. The small
   vertical full lines indicate the Fermi level for each defect
   configuration.}
 \label{PDOS}
\end{center}
\end{figure}

\begin{figure*}[t!]
  \centering
  \begin{tabular}{ccc}
  (a) bulk & (b) N$_\textrm{As}$ & (c) As$_\textrm{Ga}$
    \\ \includegraphics[width=0.6\columnwidth]{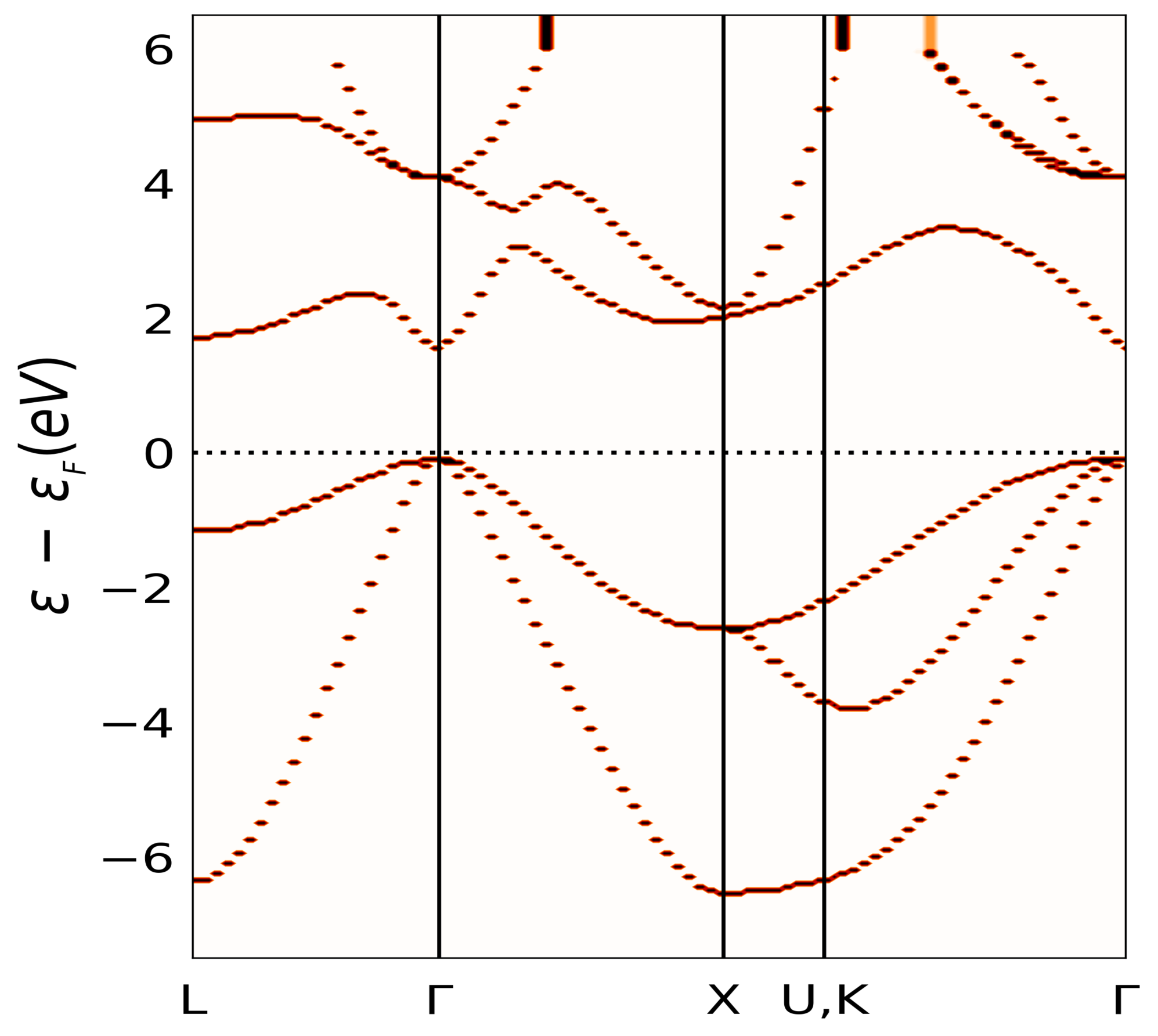} &
    \includegraphics[width=0.6\columnwidth]{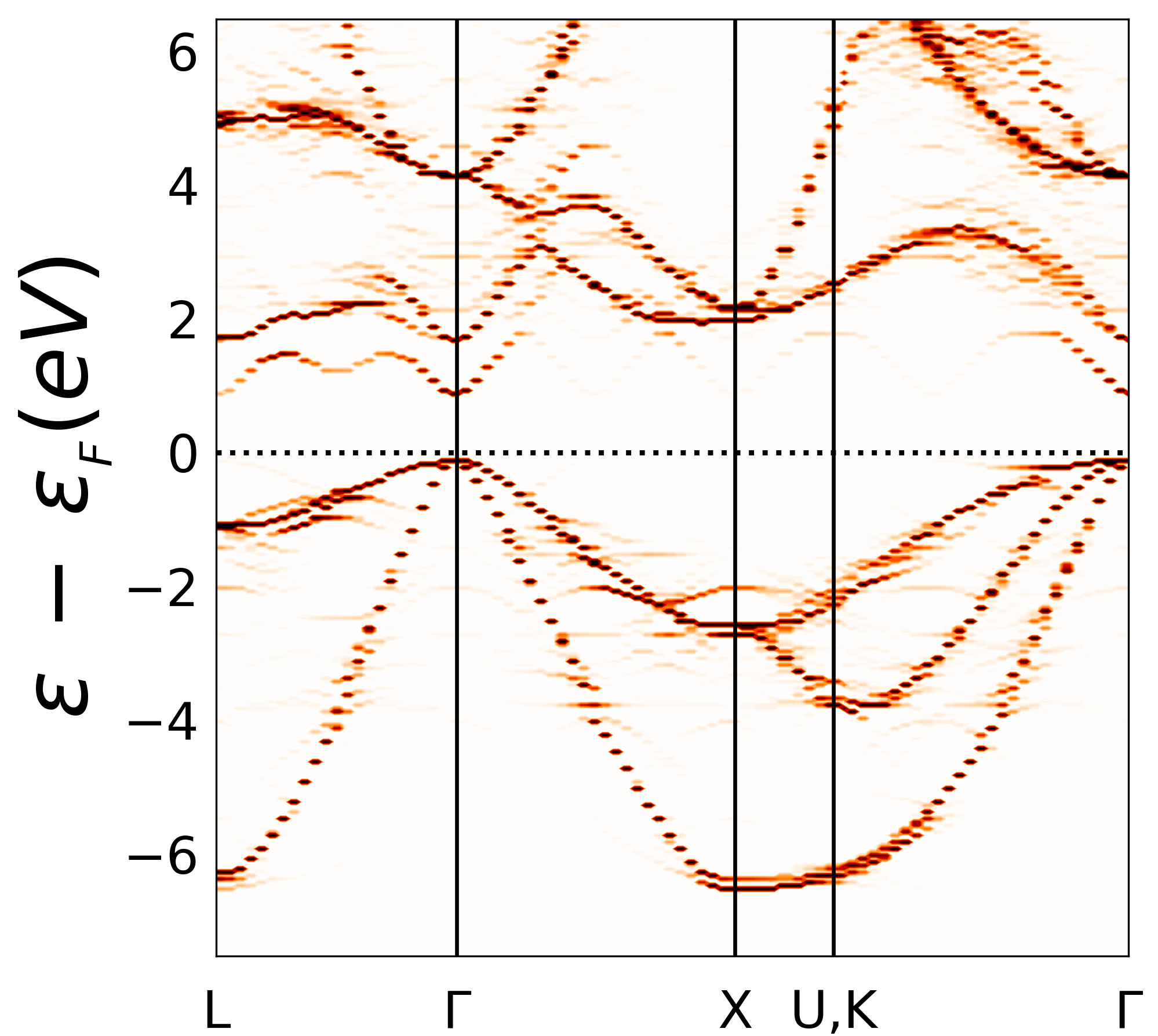} &
    \includegraphics[width=0.6\columnwidth]{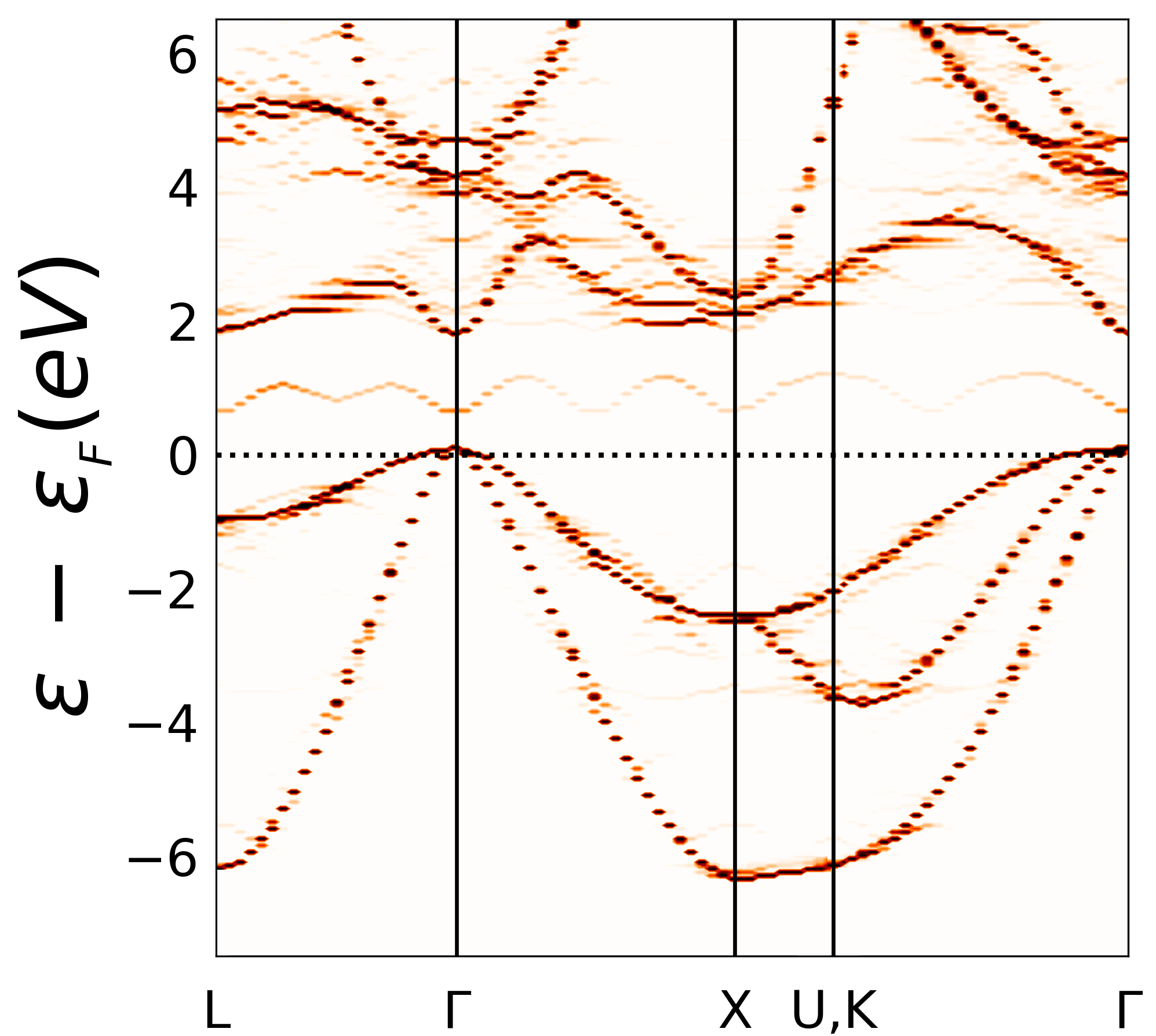}
    \\ (d) (N-N)$_\textrm{As}$ & (e) (N-As)$_\textrm{As}$ &
   (f) As$_\textrm{Ga}$-N$_\textrm{As}$ \\
    \includegraphics[width=0.6\columnwidth]{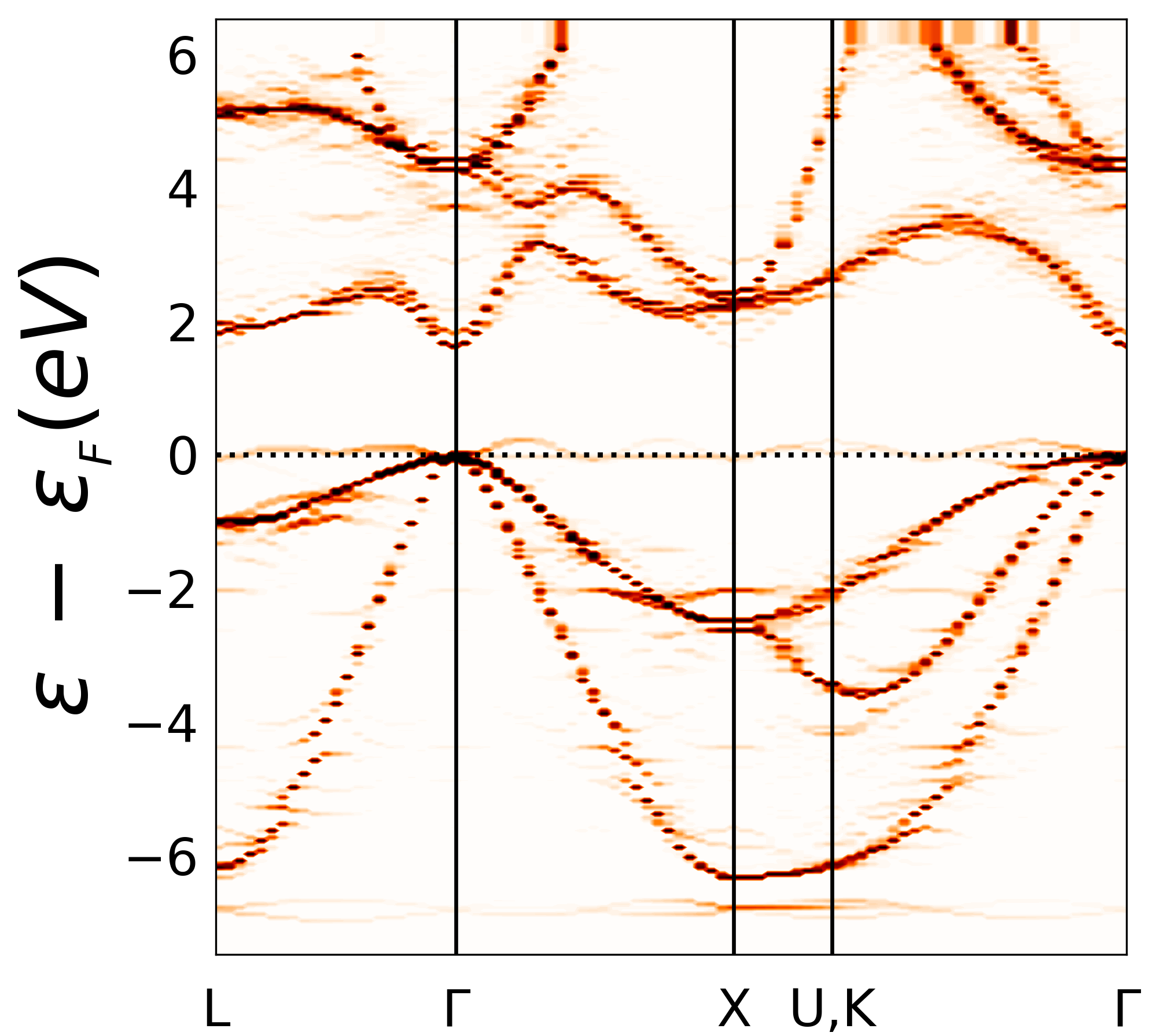} &
    \includegraphics[width=0.6\columnwidth]{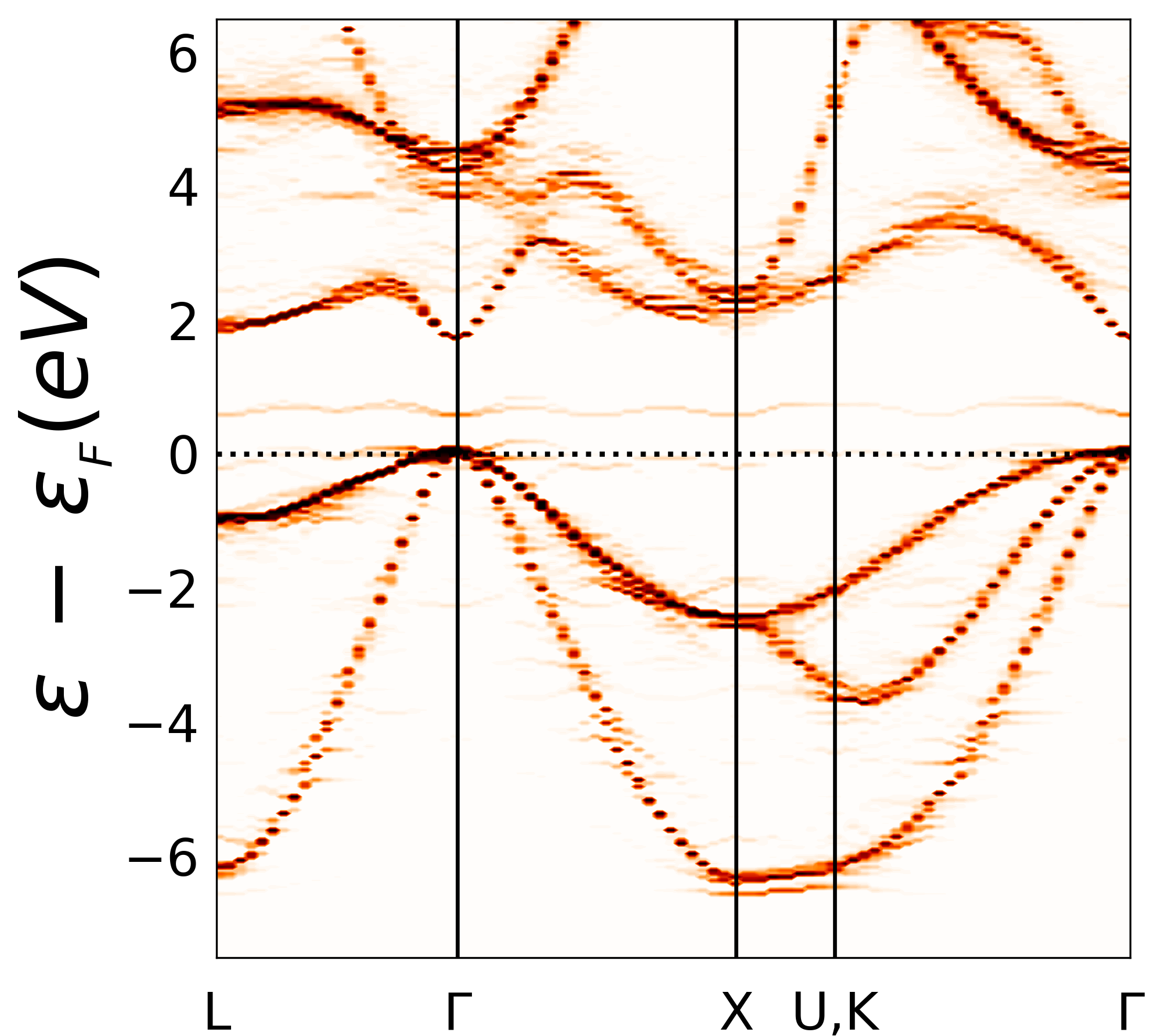} &
    \includegraphics[width=0.6\columnwidth]{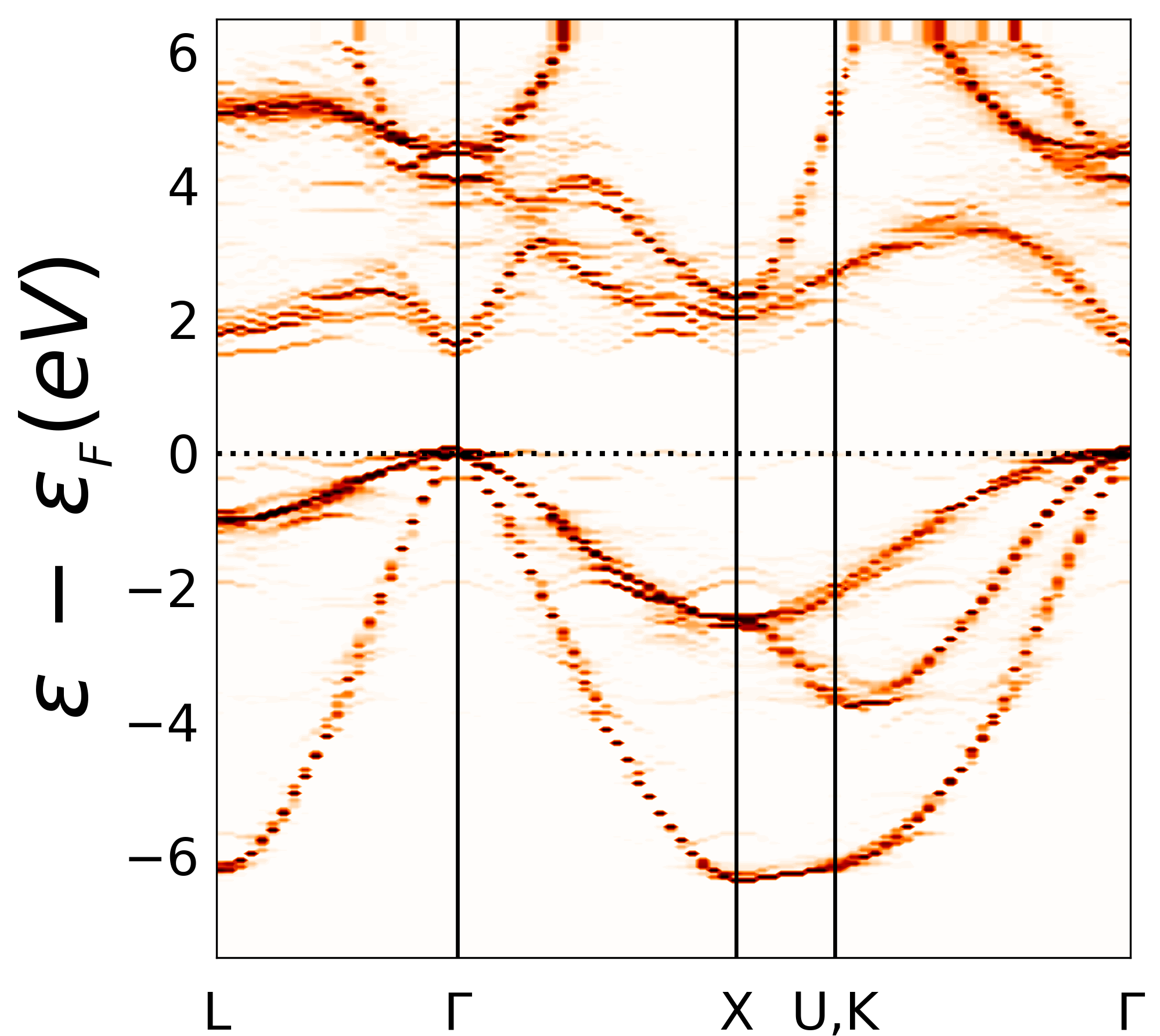}
  \end{tabular}
  \begin{center}
    \caption[]{ Effective band structure of bulk GaAs, and the five
      defect configurations unfolded into a primitive cell of the
      Brillouin zone using the BandUP
      code.\cite{madeiros2014,madeiros2015} (a) Bulk GaAs. (b) and
      (c): N$_\textrm{As}$ and As$_\textrm{Ga}$ substitutional
      defects. (d), (e) and (f): (N-N)$_\textrm{As}$,
      (N-As)$_\textrm{As}$ and As$_\textrm{Ga}$-N$_\textrm{As}$
      complex defects respectively. Here, the energy axis is referenced with respect the Fermi level of   bulk GaAs.}
    \label{unfold}
  \end{center}
\end{figure*}

In Figure~\ref{PDOS} we show the calculated total and projected
densities of states of bulk GaAs and of the five defects in
GaAs$_{1-x}$N$_{x}$ considered in this study. The projected density of
states includes the contribution from the first neighbours around each
defect configuration. The Fermi levels in the density of states in Figure \ref{PDOS} were aligned with respect to the position of a core level, and  showed with respect to 
the Fermi level of bulk GaAs. From the density of states we can get
information about the carrier type induced by each kind of defects. In
particular, we find that for the defects As$_\textrm{Ga}$,
(N-As)$_\textrm{As}$, (N-N)$_\textrm{As}$,
As$_\textrm{Ga}$-N$_\textrm{As}$, localized states can be identified
as narrow peaks within the bulk GaAs bandgap (indicated by vertical
dotted lines in Fig.\ref{PDOS}). The localized character of the
(N-N)$_\textrm{As}$ dimers was suggested from first-principles
pseudopotential method studies in
Ref. \onlinecite{Lowther2001}. Meanwhile, for substitutional N$_{As}$
no localized states within the band gap can be identified, indicating
a strong hybridization between N-states and GaAs states at the edge of
the valence and/or conduction bands.

\subsection{Unfolding of the supercell
  bandstructures}\label{unfolding}

To get meaningful information from the massive number of bands of a
GaAs with N defect systems, we employ the technique of band unfolding
by using the recently released BandUP
code.\cite{madeiros2014,madeiros2015} With this technique we can
obtain the unfolding of the bands in the Brillouin zone (BZ) of the
large supercell back into the BZ of the primitive unit cell of
GaAs.  As done for the density of states,  the Fermi levels in the density of states in Figures \ref{unfold} were aligned with respect to the position of a core level, and  showed with respect to 
the Fermi level of bulk GaAs.

As reference, the unfolded band structure of GaAs close to the bandgap
is shown in Figure~\ref{unfold}(a). The unfolded band structure for
GaAs with each of the different N defects considered in this work,
along high-symmetry directions of the primitive cell BZ are shown in
Figure~\ref{unfold}(b-d). As a general result, it can be seen that by
introducing N defects in GaAs, some new bands with low Bloch spectral
weight appear, mostly close to the bandgap.  In the following we will
discuss in detail the effect of each of the defects on the electronic
band structure of GaAs.

In Figure~\ref{unfold}(b), we exhibit the unfolded band structure of
GaAs with substitutional N$_\textrm{As}$ defects. We observe that N$_\textrm{As}$
defects mostly affect the conduction band of GaAs (compare with
Fig.~\ref{unfold}(a)), causing the formation of quasilocalized
electron states associated with the substitutional N atoms, which
interact strongly with the GaAs conduction band edge. In particular,
the bottom of the conduction band is split, and a second minimum
appears at $\Gamma$ which possesses low spectral weight. Our results
are in agreement with previous studies by Shan et al,\cite{shan1999}
where a simple physical understanding of the dramatic effect of
N$_\textrm{As}$ defects on the bandgap of GaAs was based on the band
anticrossing model (BAC), and they showed that the reduction of the energy
bandgap could be described by an interaction between the conduction
band and a higher-lying set of localized N resonant
states.\cite{lindsay2001,fahy2004}  In order to validate our results in Figure~\ref{unfold}(b), we compare our obtained band gap with the dilute limit result given by the  band anticrossing model 
in Ref. \onlinecite{walukiewicz2002}: the nitrogen concentration in our calculation 
would correspond to $\sim$1.56\% and the direct gap at is $\Gamma$ $\sim$1.108 eV, which compares quite well with the value of $\sim$ 1.07 eV given by the BAC.   Our results also  agree with our DOS
calculations presented in Fig.\ref{PDOS}, where localized N-induced
states are not identified when N is incorporated substitutionally.

Figure \ref{unfold}(c) shows the unfolded band structure of GaAs with
As$_{Ga}$ substitutional defects.  As can be seen in the figure, an
electronic state with low dispersion appears in the GaAs gap, closer
to the bottom of the conduction band, which is correlated to the sharp
peak that can be seen in the corresponding DOS (Fig.\ref{PDOS}). In
addition, we found that the As$_\textrm{Ga}$ defect does not affect much the
band edges in GaAs, opposite to the effect of the N$_{As}$
substitutional defect.

In Figure~\ref{unfold}(d) and (e), we exhibit the unfolded band
structure of GaAs with (N-N)$_\textrm{As}$ and (N-As)$_\textrm{As}$ complex
interstitial defects respectively.  As can be seen, the bottom of the
conduction band is less affected by these defects in contrast to our
findings for the susbtitutional N$_\textrm{As}$ defect.  In agreement with
results previously presented in Figure~\ref{PDOS}, (N-N)$_\textrm{As}$ induces
quite localized states in the top of the GaAs valence band, which can
be seen in the band structure as flat lines with very low spectral
weight just below the Fermi level. On the other hand, (N-As)$_\textrm{As}$
induces two localized states, one at the top of the valence band and
the other deeper in the GaAs band gap.

Finally, in Figure~\ref{unfold}(f) we exhibit the unfolded band
structure of GaAs with the As$_\textrm{Ga}$-N$_\textrm{As}$ defect. We observe that
this defect affects both the conduction and valence bands of GaAs,
producing firstly a reduction of the direct gap, as for N$_{As}$, and
secondly the formation of quasilocalized electron states near the top
of the GaAs valence bands. The lines appearing very close to the top
of the valence band in Fig.~\ref{unfold}(f) can be observed as a sharp
peak in the DOS (Fig.\ref{PDOS}).

Finally, it is worth mentioning that for devices that rely on good carrier mobility,
the process of trapping, de-trapping and scattering by the potential fluctuations
reduces mobility and hence performance.  So, localized states can result in relatively short carrier lifetimes.
Short carrier lifetimes will clearly inhibit the performance of devices such as lasers, which require long carrier lifetimes.
However, localization can also be exploited to enhance device performance. In the case of III-nitrides materials with high densities of non-radiative centres, exciton localisation improves luminescence efficiency by preventing migration of carriers towards the defects. In particular, dilute nitrides could be a good candidate material for avalanche photodetectors\cite{tan2013}, where the degradation in electron mobility actually becomes advantageous.


\section{Summary and conclusions}\label{conclusions}

In this work, we have presented calculations of the structural and
electronic properties of dilute GaAs$_{1-x}$N$_{x}$ alloys using a
supercell approach, within the framework of density functional theory
and the GGA aproximation, using a hybrid functional for the exchange
correlation functional. Motivated by the fact that semilocal DFT
functionals, like GGA, underestimate the band gap of
GaAs$_{1-x}$N$_{x}$ alloys, and that this error affects the position
of defect levels within the band gap and the values of formation
energies, we adopted the Freysoldt et al.\cite{freysoldt2016} method:
i.e., using formation energies and defect levels as calculated with
GGA, but interpreting them within a bandgap where the valence band
edge has been shifted down, and the conduction-band edge has been
shifted up, as obtained from a HSE calculation.

We studied the impact of five different N defects on the electronic
structure of GaAsN alloys: we calculate the electronic states,
formation energies and charge transition levels. The set of studied
defects includes N$_\textrm{As}$, As$_\textrm{Ga}$,
As$_\textrm{Ga}$-N$_\textrm{As}$ substitutional defects, and
(N-N)$_\textrm{As}$, (N-As)$_\textrm{As}$ split-interstitial complex
defects. We find that the formation energy of the neutral
As$_\textrm{Ga}$ defect is lower than that of substitutional
N$_\textrm{As}$. Meanwhile, the formation energy of neutral
(N-N)$_\textrm{As}$ is higher than for all the other defects, while
our results also indicate that the (N-As)$_\textrm{As}$ split
interstitial is the dominant interstitial complex defect in the
neutral state of dilute GaAsN alloys. These results are in agreement
with recent experimental results obtained using Rutherford
backscattering spectroscopy and nuclear reaction analysis spectra with
Monte Carlo simulations,\cite{jen2015} and other reports for GaAsN and
related dilute nitride
alloys.\cite{ishikawa2011,kim2013,chen2010,pham2007,hari2012,buckeridge2014}
As a prediction, we find that the antisite As$_{Ga}$ defect is the
most favorable one in GaAs$_{1-x}$N$_{x}$.  The RBS spectra measured
in Ref.\onlinecite{jen2015} suggested the possible presence of Ga
interstitial and/or As antisite in the GaAs$_{1-x}$N$_{x}$ alloys.

For the supercell calculations, the band structures were unfolded.
Although nitrogen is isoelectronic with arsenic, because of the large
difference in size and electronegativity between N and As atoms, this
causes the formation of quasilocalized electron states associated with
the substitution of N atoms, which interact strongly with the GaAs
conduction band, and leads to a substantial reduction of the bandgap.
We find that the largest changes in the band structure are produced by
an isolated N atom in GaAs, which is resonant with the conduction
band, exhibiting a strong hybridization between N and GaAs
states. Deeper levels in the energy gap, for which the energy required
to transfer an electron (or hole) towards the conduction (or valence)
band exceeds the characteristic thermal energy ($k_{B}T$), are
obtained with (N-N)$_\textrm{As}$ and (N-As)$_\textrm{As}$ split-interstitial
defects. Our results agree with those given by the band anticrossing model, which is appropiate to describe the electronic structure of GaAsN in the dilute limit.\cite{walukiewicz2002}

\section{Acknowledgments}

C.I.V. and J.D.F. are Investigadores Cient{\'{\i}}ficos of CONICET
(Argentina).  J.D.Q.-F. thanks CONICET for a postdoctoral fellowship
and acknowledges support of PIP 0650 (CONICET) grant. C.I.V.
acknowledges support from CONICET (PIP-2015-11220150100538CO) and
ANPCyT (PICT-2012-1069) grants. The authors would like to thank
Dr. Paulo V. C. Medeiros and his co-workers for their BandUP code. We
thank the Centro de Simulaci{\'o}n Computacional p/Aplicaciones
Tecnol{\'o}gicas (CSC-CONICET) for granting the use of computational
resources which allowed us to perform much of the simulations included
in this work.

\bibliographystyle{apsrev4-1.bst} 
\bibliography{references2} 

\end{document}